\documentclass[prl,twocolumn,floatfix,superscriptaddress,longtable,notitlepage,10pt,nofootinbib]{revtex4-2}
\usepackage{amsmath}
\usepackage{lipsum} 
\usepackage{verbatim}
\usepackage{epsfig}
\usepackage{subfigure}
\usepackage{graphicx}
\usepackage{enumitem}
\usepackage{amsfonts}
\usepackage[final]{showlabels}
\usepackage{enumitem}
\usepackage[figuresright]{rotating}
\usepackage{amssymb}
\usepackage{amsmath}
\usepackage{psfrag}
\usepackage{mathtools}
\usepackage[export]{adjustbox}
\usepackage{bm}
\usepackage[colorlinks,linkcolor=blue,anchorcolor=blue,citecolor=blue,urlcolor=blue]{hyperref}
\usepackage[version=4]{mhchem}
\usepackage[svgnames]{xcolor}

\def\be{\begin{equation}} \def\ee{\end{equation}}
\def\bea{\begin{eqnarray}} \def\eea{\end{eqnarray}}

\def\bpm{\begin{pmatrix}} \def\epm{\end{pmatrix}}

\makeatletter
\newcommand*{\balancecolsandclearpage}{%
	\close@column@grid
	\clearpage
}
\makeatother

\begin{document}
	\title{Planckian bound on quantum dynamical entropy} 
	\author{Xiangyu Cao}
	\affiliation{Laboratoire de Physique de l'\'Ecole normale sup\'erieure, ENS, Universit\'e PSL, CNRS, Sorbonne Universit\'e, Universit\'e Paris Cit\'e, F-75005 Paris, France}
\date{\today}
\begin{abstract}
We introduce a simplified version of Connes-Narnhofer-Thirring's quantum dynamical entropy for quantum systems. It quantifies the amount of information gained about the initial condition from continuously monitoring an observable. A nonzero entropy growth rate can be obtained by monitoring the thermal fluctuation of an extensive observable in a generic many-body system, away from classical or large $N$ limits. We explicitly compute the entropy rate in the thermodynamic and long-time limit, in terms of the two-point correlation functions. We conjecture a universal Planckian bound for the entropy rate. Related results on the purification rate are also obtained. 
\end{abstract}
	\maketitle
	
 \noindent\textbf{Introduction}. The dynamical entropy of Kolmogorov and Sinai (KS)~\cite{kolmogorov,sinai1959,sinai2009} is an important concept in classical mechanics and thermodynamics. It can be defined as the amount of information that an observer may gain about the initial condition by continuously monitoring the system. The dynamical entropy is a fundamental diagnosis of classical chaos~\cite{walters2000introduction}. The butterfly effect amplifies arbitrarily small details of the initial condition, and makes them observable at late time. This leads to a nonzero dynamical entropy rate, which is quantitatively tied to the Lyapunov exponents by Pesin's relation~\cite{pesin2008}. Further relations have been established with thermodynamic entropy production~\cite{latora-baranger} and transport~\cite{latora-baranger,gaspard-nicolis}.

 Defining a dynamical entropy for quantum systems has been an outstanding question since KS~\cite{gutzwillerbookchaos}. A few proposals have been put forward~\cite{connes-stormer,CNT,voiculescu-entropy,AF,slom-entropy,Hudetz1998,accadi-entropy,benatti2012deterministic,Jorge}, and concrete calculations have been performed for simple systems such as free quantum gases~\cite{narnfoger-thirring,Hudetz1998,benatti98}. However, it remains unknown whether quantum dynamical entropy can characterize chaos and complexity in many-body quantum systems (see \cite{Prosen_2007} for a recent survey). This question is timely as many-body quantum chaos has been an intensively pursued topic~\cite{otoc-ruc,otoc-ruc-pollman,prosen-mbqc,deluca,RUC-review,D'Alessio-review,circuit-complexity,susskind-complexity,Sekino_2008,zoller-chaos,deep-therm,uogh,fleishbauer-simulation,bertini-op-entangle,dubail-op-entangle,lai-web}.  {One interesting approach is the out-of-time order correlation (OTOC)~\cite{bound-chaos},
 \begin{equation} \label{eq:OTOC}
 	\text{OTOC} =	\mathrm{Tr}[ \sqrt\rho [P, Q(t)]^\dagger \sqrt\rho [P, Q(t)] ]
 \end{equation}
 where $\rho =  e^{-\beta H} / \mathrm{Tr}[e^{-\beta H}]$ is the thermal state at temperature $1/\beta$ of some quantum system with Hamiltonian $H$, $P$ and $Q$ are operators, and $Q(t) = e^{i H t} Q e^{- i H t}$. OTOCs generalize the classical butterfly effect, that is, the sensitivity of  an observable at $t$ to initial perturbations, to quantum systems. Remarkably, its growth rate (Lyapunov exponent) obeys a Planckian bound at low temperatures~\cite{bound-chaos},
 \begin{equation} \label{eq:OTOCbound}
 	\text{OTOC} \sim e^{\lambda_L t} \implies \lambda_L \le 2 \pi T, T = 1/\beta.
 \end{equation}
The Planckian bound showcases a universal limit on dynamical chaos due to quantum effects, and is famously saturated by strongly interacting holographic models~\cite{sachdevye,kitaev,syk-comment,kitaev-suh,syk-review}; see also \cite{srednicki-bound,delacretaz-bound,review-planckian,bound-luca,bound-viscosity,bounds-nussinov} for other ``Planckian bounds''. However, OTOCs do not have a well-defined exponential growth in \textit{generic} many-body quantum systems with local interactions~\cite{otoc-ruc-pollman,khmeni-lyapu,sarang-otoc-int}, that is, away from semiclassical or large $N$ limits. Meanwhile, quantum chaos diagnosis inspired by random matrix theory~\cite{casati,bgs,berry-tabor} probes time scales that diverge in the thermodynamic limit. 
 	
 In this Letter, {we propose a quantum chaos probe that remedies the above issues. It is a simplified version of  the Connes-Narnhofer-Thirring (CNT) quantum dynamical entropy~\cite{CNT}. Like the OTOC, it is defined for a thermal state $\rho$ and an operator $Q$. Unlike OTOC, it involves repeatedly measuring (monitoring~\cite{benatti98,Gaspard1992-qchaos,deutsch-tomography-chaos}) $Q$ from $0$ to $t$, and is defined as follows, 
 \begin{equation} \label{eq:SCNT-intro}
 S_{\text{CNT}}	 := S_{\text{cl}}(p) - \sum_{s \le t} \left[ S_{\text{cl}}(p^s) - J_s  \right] \,,
 \end{equation}
 where $S_{\text{cl}}(p)$ is the classical Shannon entropy of the outcome distribution, $ S_{\text{cl}}(p^s)$ that of the marginal distribution at time $s$, and $J_s$ the amount of information gained by the measurement at $s$; see below and in particular \eqref{eq:rhop-marginal}-\eqref{eq:SCNT} for more details. { As we shall see,} The CNT entropy may be \textit{more} accessible in experiments than recently studied measurement-induced quantities~\cite{noel-MIPT,MIPT-exp1}.
 
Remarkably, for suitable $Q$ and measurement scheme, $S_{\text{CNT}}$ has a well-defined linear growth, in \textit{generic} quantum many-body systems, and at a time scale independent of the system size (and only determined by the local energy scale). Moreover, we conjecture that the growth rate satisfies a Planckian bound:}
\begin{equation} \label{eq:boundintro}
	\lim_{t \to\infty }	\lim_{V\to\infty} S_{\text{CNT}} / t \le c T \,,
\end{equation}
where $V$ is the volume, and $c$ is a universal constant. 

We support the conjecture \eqref{eq:boundintro} with two main pieces of evidence. First, its growth rate vanishes if $Q$ is a local observable. Second, if $Q$ is a rescaled extensive quantity the entropy growth rate is given, in a large non-critical many-body systems and for a Gaussian measurement scheme, by an exact formula, which satisfies a Planckian bound. The exact rate formula is obtained thanks to a general emergent Gaussianity property of interacting systems, and is our main technical contribution. We shall then argue that the Planck bound applies more generally. 

 Below, we shall define our setup in more detail, see Fig.~\ref{fig:setup}, go through the arguments supporting the Planckian bound, and discuss analogous results on purification.

%


\begin{figure}
	\centering
	\includegraphics[width=.5\columnwidth]{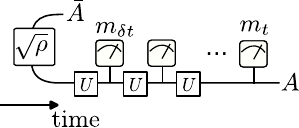}
	\caption{ The purification setup. Two identical systems $A\bar{A}$ are initialized in an thermal field double state $\sqrt{\rho}, \rho =  e^{-\beta H} / \mathrm{Tr}[e^{-\beta H}]$. $A$ is evolved by $U = e^{-i t H}$ and repeatedly measured, yielding outcomes $m_{\delta t}, \dots, m_t$. The latter disentangles (purifies) $\bar{A}$, revealing information about the initial condition. }
	\label{fig:setup}
\end{figure}
 \noindent\textbf{Setup.}   
{ Consider a many-body quantum system $A$ with short-range Hamiltonian $H$, initialized in a thermal state $\rho = e^{-\beta H} / \mathrm{Tr}[e^{-\beta H}]$ at temperature $1/\beta$. We purify $\rho$ as the reduced density matrix of the pure state $\sqrt{\rho}$ on $A \bar{A}$, where $\bar{A}$ is an identical copy of $A$. The state $\sqrt\rho$ is known as the thermofield double (TFD), see \cite{tsie-thermaldouble,chen2023quantumprep} for realization proposals on near-term quantum platforms.} 
	
The system $A$ undergoes time evolution under $H$, whereas $\bar{A}$ is left intact.  In addition, we monitor an operator $Q$ acting on $A$ at $t = \delta t, 2 \delta t, \dots$. The measurement is described by a family of Kraus operators $\{ K_m \}_m$ depending on $Q$, indexed by possible measurement outcomes $m$, and satisfying unitary $\sum_m K_m^\dagger K_m = \mathbf{1}$~\cite{Wiseman_Milburn_2009}. For example, we may take
$$K_{m} = ( \mathbf{1}+ m Q) / \sqrt{2 ( \mathbf{1} + m^2 Q^2)}, m = \pm 1, $$
or \eqref{eq:Km} below. Doing this until $t = n \delta t$, we obtain a random sequence of outcomes $\vec{m} = (m_{\delta t}, m_{2 \delta t}, \dots, m_t)$. In terms of
\begin{equation}
	\tilde{\rho}_{\vec{m}}  :=  \sqrt{\rho}    \prod_{s = t}^{\delta t} K_{m_s}^\dagger(s)    \prod_{s = \delta t}^t K_{m_s}(s)  \sqrt{\rho}   \,,\,  \label{eq:rhomtilde}
\end{equation}
where $O(t) := e^{i H t} O e^{-i H t}  $, the outcome probability and the post-measure state of $\bar{A}$ are respectively
\begin{equation}
	p_{\vec{m}} = \mathrm{Tr}[	\tilde{\rho}_{\vec{m}} ] \,,\, 
	\rho_{\vec{m}}  = 	\tilde{\rho}_{\vec{m}}  / p_{\vec{m}}  \,. \label{eq:rhom}
\end{equation}
 In terms of these, purification is defined as
\begin{equation} \label{eq:J}
	J := \sum_{\vec{m}} p_{\vec{m}}  \mathrm{Tr}[\rho_{\vec{m}}  \ln \rho_{\vec{m}}  - \rho  \ln \rho ]   \,,
\end{equation} 
{Namely, it equals how much the continuous monitoring reduces the entanglement between $A$ and $\bar{A}$. $J$ quantifies the amount of information (on the initial condition) revealed by the measurements~\cite{Henderson_2001,discord-zurek}, and has been used as a probe of measurement-induced phases~\cite{gullans-huse-prx,skinner19,li-fisher,bao-altman,vasseur-ludwig}.  }

For the CNT entropy, we introduce, for each $s = \delta t, 2 \delta t, \dots, t$,  the marginal analogues of \eqref{eq:rhomtilde}, \eqref{eq:rhom}, and \eqref{eq:J},
\begin{align}
&	\tilde{\rho}^s_{m} := \sum_{\vec{m}: m_s = m} \tilde{\rho}_{\vec{m}}  \,,\, p^s_{m} := \mathrm{Tr}[\tilde{\rho}^s_{m}] \,,\, \rho^s_{m} := \frac{\tilde{\rho}^s_{m} }{p^s_{m}} \,, \label{eq:rhop-marginal}  \\ 
&	J_s :=   \sum_{m} p_{m}^s \mathrm{Tr}[\rho^s_{m} \ln \rho_{m}^s - \rho  \ln \rho ]   \,.  \label{eq:Js-def}
\end{align}
{Then the CNT entropy's definition \eqref{eq:SCNT-intro} is completed by the recalling the definition of Shannon entropies,
\begin{align} \label{eq:SCNT}
		S_{\text{cl}}(p) :=  \sum_{\vec{m}}  -p_{\vec{m}}  \ln p_{\vec{m}} , 
		S_{\text{cl}}(p^s)  :=   \sum_{{m}}  - p^s_{{m}}  \ln p^s_{{m}} .
 \end{align}  
The CNT entropy is in principle accessible experimentally. $S_{\text{cl}}(p) $ is the Shannon entropy of a random time sequence (sampled by experimental runs), and can be estimated by classical statistical methods~\cite{entropy-estimate}.  $S_{\text{cl}}(p^s) $ concerns 1D distributions, so is simpler to estimate. In $J_s$, the entanglement entropies can be estimated by modern tomography techniques~\cite{tomography}. Crucially, there is not a hard post-selection problem: Since we condition on a single outcome $m_s$, we do not need $O(e^{t})$ number of experimental runs to achieve one outcome many times in order to perform tomography. Purification is more challenging, as it involves an $O(e^t)$ hard post-selection~\cite{noel-MIPT,MIPT-exp1,garratt-post}. }

Our choice to monitor a single observable $Q$ is motivated by the comparison with OTOC, and ultimately by the Planckian bound, as the entropy can be trivially increased by monitoring multiple uncorrelated regions. This makes our definition simpler than CNT's proposal, which involves maximizing over measurement schemes.

{ Nevertheless, our definition retains a key ingredient of CNT's work, namely the term $S_{\text{cl}}(p^s)  - J_s $ in \eqref{eq:SCNT}, known as the ``entropy defect''.} It accounts for a feature of quantum mechanics: Measurements backreact on the system, and can thus prevent later observations from gaining information about the initial state~\cite{benatti98,benatti2012deterministic}. For this reason, $S_{\text{cl}}(p)$ alone is not a good definition of quantum dynamical entropy. Also, { the growth of $S_{\text{CNT}}$ depends qualitatively on the nature of the operator $Q$, as we discuss next.}



 \noindent \textbf{Local operators}.  We first consider the case where $Q$ is a local operator acting on $O(1)$ local degrees of freedom. We assume those to be non-semiclassical, and have a small Hilbert space.  We argue that the asymptotic CNT entropy rate cannot be positive for a generic dynamics, due to the significant backreaction (dephasing).

In fact, we argue that $J_t \to 0$ as $t \to \infty$; from the definition \eqref{eq:SCNT-intro} and the subaddictivity of the Shannon entropy, we have $ S_{\text{CNT}} / t \le \sum_{s\le t} J_s / t$, so $J_t \to 0$ would forbid a positive entropy rate. For that, observe from \eqref{eq:rhop-marginal}, \eqref{eq:rhomtilde}, and unitarity, that 
\begin{equation} \label{eq:rhot-Heisenberg}
	\tilde\rho^{t}_{m} =  \sqrt\rho O_m   \sqrt\rho  \,,\, O_m =  \mathcal{L}  ( \mathcal{D} \mathcal{L})^{n-1} [K_m^\dagger K_m]  \,.
\end{equation}
Here $n = t / \delta t $ and
\begin{equation} 
\mathcal{L}[O] := e^{i H t} O e^{-i H t} \,,\,	\mathcal{D}[O] := \sum_m K_m^\dagger O K_m \,,\, 
\end{equation}
are the time-evolution and dephasing channel (induced by a measurement with discarded result). For concreteness,  consider strongly measuring a spin in a lattice along the $z$ direction, then $K_{m = \pm} = (1 \pm Z) / \sqrt{2}$ ($Z$ is the Pauli acting on the spin), and $\mathcal{D} [(a X + b Y) O'] = 0$ for any $a, b$, where $X$ or $Y$ acts on the monitored spin and $O'$ is any operator acting elsewhere. Such terms are repeatedly generated by a generic enough $\mathcal{L}$ and annihilated by $ \mathcal{D} $, until the operator $ ( \mathcal{D} \mathcal{L})^n [K_m^\dagger K_m] \to c \mathbf{1} $ as $n \to \infty$. Thus, $ \tilde\rho^{t}_{m}  $ tends to be  $\propto \rho$ and $J_t \to 0$. 

\begin{figure}
	\centering
	\includegraphics[width=1.\columnwidth]{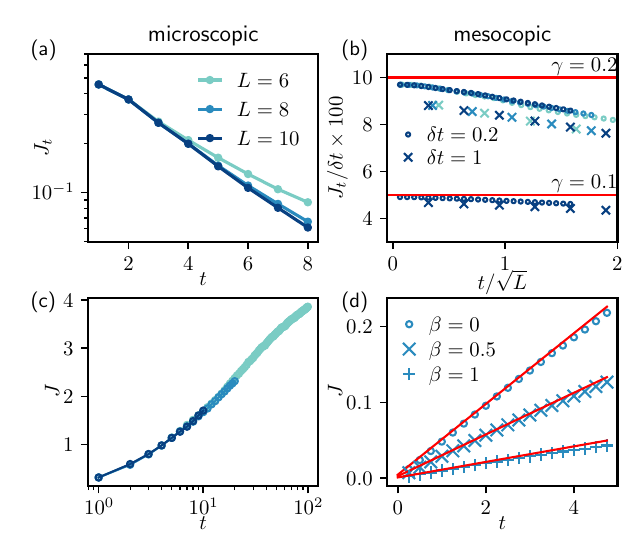}
	\caption{Numerical calculation of the quantity $J_s$ ~\eqref{eq:Js-def} and purification $J$~\eqref{eq:J} with micro and mesoscopic monitoring, on the mixed-field Ising chain $H = \sum_{j=1}^L X_j + h Z_j + J Z_j Z_{j+1}$ ($L + 1 \equiv 1$), $h = 1.245, J = 0.945$, with $L = 6, 8, 10$ [with color code in panel (a)], at $\beta = 0$ by default. \textbf{(a)}$J_s$ with microscopic monitoring, $\delta t = 1$,  with Kraus operators $K_{m = \pm} = \sqrt{(\mathbf{1} \pm O)/2 }$ with $O = 0.75 Z_0 Z_2$. \textbf{(b)} $J_s$ with mesoscopic monitoring \eqref{eq:Km}, with $\delta t = 0.2, 1$, $Q = \sqrt{\gamma / L} \sum_j Z_j$, $\gamma = 0.1, 0.2$. The horizontal lines show the analytical prediction. The time is rescaled by $1/\sqrt{L}$. \textbf{(c)} Purification $J$ with the same microscopic monitoring scheme as in panel (a). \textbf{(d)} Purification $J$ with mesoscopic monitoring, $Q = \sqrt{\gamma / L} \sum_j (Y_j - \left< Y_j \right>)$ $\gamma = 0.1$, $\delta t = 0.25$, $\beta = 0, 0.5, 1$. The straight lines represent the analytical prediction~\eqref{eq:jmeso}. See~\cite{supp} for numerical methods. 
	} \label{fig:num}
\end{figure}
The above argument can be adapted to any weak measurements on a few sites, as the dissipative evolution $\mathcal{D L}$ will only preserve the identity operator \textit{generically}\footnote{An exceptional case is when $Q$ is conserved by the dynamics. In that case, $J_t \not\to 0$. However, since $Q(t) = Q$, the outcome Shannon entropy cannot grow linearly, so the dynamical entropy rate still vanishes. Another fine-tuned counter-example is spatial translation~\cite{CNT} where the dynamical entropy rate is nonzero.}.  Numerically we observe that $J_t$ decays exponentially in a non-integrable spin chain, see Fig.~\ref{fig:num}-(a). 


 \noindent \textbf{Mesoscopic operators}. As local operators cannot yield a positive entropy rate, we are led to consider measuring a ``\textit{mesoscopic} observable'' of the form 
\begin{equation} \label{eq:defmeso}
	Q = \frac{1}{\sqrt{V}} \sum_x Q_x \,,\, \left< Q_x \right> = 0 \,,
\end{equation}
where $ \left< A \right> := \mathrm{Tr}[\rho A]$,  $Q_x$ is a local operator centered around $x$, and $V$ is the volume (number of sites in a lattice model). We may always assume $ \left< Q_x \right> = 0 $ upon replacing $Q_x \to Q_x - \left< Q_x \right> $, so that $Q$ describes the fluctuation of an extensive quantity around its average (hence the term ``mesoscopic'').  We consider the family of Kraus operators indexed by a continuous outcome
\begin{equation} \label{eq:Km}
	K_m = \frac1{(2\pi)^{\frac14}} e^{- \left(Q \sqrt{\delta t} - m \right)^2 / 4} \,,\,   m \in \mathbb{R} \,,
\end{equation}
which describe a weak measurement of $Q$ with order-one precision. The definition~\eqref{eq:SCNT} can be adapted by replacing sums by integrals; $p_{\vec{m}}$ and $p_m^s$ become probability densities and the Shannon entropies become differential entropies. The measurement strength can be tuned by rescaling $Q$ itself. The $\sqrt{\delta t}$ factor allows a well-defined continuous monitoring $\delta t \to 0$ limit.  Such mesoscopic monitoring has a qualitatively milder backreaction, and can lead to a nonzero entropy growth rate.

\noindent\textbf{Rate formula}. In fact, the aforementioned rate can be analytically calculated, even in interacting systems. Indeed, in the thermodynamic limit and away from thermal criticality, the time-dependent fluctuation of $Q$ becomes effectively Gaussian due to cluster decomposition (see End Matter). Thus, the dynamical entropy rate can be computed solely in terms of the Keldysh and retarded Green functions in frequency domain~\cite{Kamenev_2011}. We define them with the following convention:
\begin{subequations}
	\begin{align}
		&G_K(\omega) = \frac12 \int_{-\infty}^{\infty}  \left<  Q Q(t) + Q(t) Q   \right> e^{i \omega t}  d t, \\
		& G_R(\omega) =  \frac{i}2  \int_{0}^{\infty}  \left<  [Q(t), Q]   \right> e^{i \omega t}    dt .
	\end{align}
\end{subequations}
At thermal equilibrium, they are related by the fluctuation-dissipation theorem (FDT), $\mathrm{Im} G_R(\omega) =  \tanh (\beta\omega) G_K(\omega) / 2 $; $\mathrm{Re} G_R(\omega) $ is given by Kramers-Kronig. In terms of the eigenstate thermalization hypothesis~\cite{srednick,Deutsch_2018,D'Alessio-review}, the Green functions encode the same data as the two-point structure factor (yet Gaussianity is not equivalent to the vanishing of higher free cumulants~\cite{ETHplus,laura-jorge}). The emergent Gaussianity enables a simple method to compute the dynamical entropy. We write a free boson model in which a displacement operator has the same $G_K$, and perform exact calculation therein. The calculation is lengthy yet elementary; see \cite{supp} for details. The result is the following:
	\begin{align} \label{eq:rateformula}
	  &s_{\text{CNT}} := \lim_{\substack{t \to \infty \\ \delta t \to 0}}  \lim_{V\to\infty} (S_{\text{CNT}} / t)   \\
	= & \frac12 \int \frac{d \omega}{2\pi} \  \left[ \ln (1 + \tilde{G}) - \tilde{G}    + \frac{\beta \omega}{\sinh \beta \omega} G_K  \right]  \nonumber 
\end{align}
where $\tilde{G}(\omega) := G_K(\omega) + |G_R(\omega)|^2$. In particular, the limit of the quantity $J_t$ is given by~\cite{garratt-mcginley},
\begin{equation} \label{eq:Jt-prediction}
\lim_{\substack{t \to \infty \\ \delta t \to 0}}  \lim_{V\to\infty} J_t  =  \frac12 \int \frac{d \omega}{2\pi} \frac{\beta \omega}{\sinh \beta \omega} G_K(\omega). 
\end{equation}
We emphasize that \eqref{eq:rateformula} and \eqref{eq:Jt-prediction} are exact for \textit{interacting} systems, not just free bosons, thanks to emergent Gaussianity. 
{ In Fig.~\ref{fig:num}-(b), we compare this prediction to numerics in a non-integrable quantum spin chain for moderate length $L$. Although the numerical value deviates from the prediction as $t$ increases for any fixed $(L, \delta t)$, we observe a single-variable scaling $ J_t(t, {L}, \delta t)= g(t / \sqrt{L}, \delta t) $ where $g(0, \delta t)$  tends to the prediction as $\delta t \to 0$. So, if we send $L \to \infty$ and then $t \to \infty, \delta t \to 0$, as in \eqref{eq:Jt-prediction}, we find a nice agreement between theory and numerics.
}
	

\noindent\textbf{Planckian bound}. The entropy rate given by \eqref{eq:rateformula} satisfies a Planckian bound. To see this, we make a change of variable $y = \beta \omega$ in its right hand side, which then equals $\beta^{-1}$ times a $\beta$-independent integral, if we view $G_K$ and $G_R$ as functions of $y$. An explicit upper bound is obtained by maximizing the integral with respect to $G_K$ and $G_R$ ignoring FDT. This gives the optimum $|G_R|^2 = 0, G_K = y / (\sinh y - y)$, and
\begin{equation} \label{eq:bound1}
	 \beta 	s_{\text{CNT}}    \le  \int \frac{dy}{4\pi} \left[ \ln \frac{\sinh (y)}{\sinh (y)-y} - \frac{y}{\sinh y} \right] \approx 0.57. 
\end{equation}
See End Matter for a tighter bound taken into account FDT, and Fig.~\ref{fig:GKoptimal} for the optical $G_K$. 

{ 
We conjecture that the Planckian bound applies to any many-body system away from thermal criticality, and for a general class of operators and measurement schemes. We just showed it for extensive operators rescaled by $1/\sqrt{V}$ and the Gaussian scheme~\eqref{eq:Km}. We showed also that more local operators in general yield a vanishing dynamical entropy rate. Changing the rescaling factor $1/\sqrt{V}$ to another power would amount to making $G_K, G_R$ very big or small in \eqref{eq:rateformula}, and that cannot break bound \eqref{eq:bound1}: The optimum has $G_K, G_R \sim O(1)$. Finally, we cannot prove, but we surmise the Gaussian scheme~\eqref{eq:Km} to be a renormalization group fixed point of a large class of schemes under temporal coarse-graining. So, all these schemes should satisfy a universal Planckian bound in the most nontrivial low-$T$ limit [Planckian bounds like \eqref{eq:OTOCbound} and \eqref{eq:bound1} are usually far from tight at high-$T$ since the growth rate is limited by the local energy scale and does not diverge with $T$.] In light of these arguments, the Planckian bound \eqref{eq:boundintro} is a rather plausible conjecture. 

We remark that in the rate formula~\eqref{eq:rateformula}, the factor $\beta \omega / \sinh (\beta \omega)$ is exponentially small for $|\omega| \gg 1/\beta$.  So, at low temperature, only slow fluctuations beyond the ``Planckian'' timescale $\tau \gtrsim \beta$ contribute to the dynamical entropy. Indeed, the information on $\bar{A}$ accessible from $A$ is encoded in correlation functions across the TFD, of the form $\mathrm{Tr}[ Q \sqrt{\rho} Q(t) \sqrt{\rho} ]$, which are known to decay as $e^{-\beta |\omega| /2}$ or faster at large $\omega$~\cite{uogh,srednicki-bound}. In this regard, our Planckian bound resembles that of OTOC, whose derivations~\cite{bound-chaos,bound-ueda,jorge-bound} often rely on the TFD in \eqref{eq:OTOC}; see however \cite{galitski-otoc-regular,romero-otoc-regular,syk-kobrin}. 
}

\noindent \textbf{Purification}. Under the same assumptions and using the same methods as \eqref{eq:rateformula}, we may obtain an exact formula for the purification rate~\cite{supp}, 
\begin{equation}\label{eq:jmeso}
\mathcal{J} :=  \lim_{\substack{t \to \infty \\ \delta t \to 0}}  \lim_{V\to\infty} \frac{J}t =  \int \frac{d\omega}{4\pi}  \frac{\beta \omega}{\sinh \beta \omega} \frac{G_K}{1 + G_K + |G_{R}|^2}.
\end{equation}
 {This result complements studies of purification at time $t \sim e^{cV}$~\cite{bulchandani-purify,deluca2023universality}.} We checked \eqref{eq:jmeso} with exact numerical simulations in small systems and find an excellent agreement, see Fig.~\ref{fig:num}-(d).  Eq.~\eqref{eq:jmeso} also satisfies a Planckian bound: Since $G_K , |G_{R}|^2 \ge 0$, we have $G_K / (1 + G_K + |G_{R}|^2 )\le 1$, and $ \beta \mathcal{J} \le \pi / 8$. For local observables, we observe numerically that $J/t \to 0$ as well; in fact, $J  \sim \ln t$ in our example, see Fig.~\ref{fig:num}-(c). So we conjecture that purification also satisfies a general Planckian bound. In summary, purification has similar properties as dynamical entropy, while being simpler to measure numerically (but not experimentally). 


 \noindent  \textbf{Conclusion}. { We proposed probes of low-energy quantum chaos that have promising behaviors for \textit{generic} quantum many-body systems away from semiclassical and large $N$ limits, and worth further theoretical or experiment study.} 
 
{Under quite general conditions, these probes depend only on linear response functions, and do not distinguish directly integrable and non-integrable systems (unlike random-matrix inspired probes~\cite{berry-tabor,bgs}). Indeed, one can reproduce any linear response function in a free system. Yet, while not probed \textit{directly}, integrability often leads to distinct linear response properties and influences dynamical entropy indirectly. It will be useful to study this point in detail.}

We close with a few other directions for future work. We would like to understand corrections to the emergent Gaussianity, and explore connection with nonlinear hydrodynamic tails~\cite{gaspard-nicolis,mukerjee-hydro,liu-eft,delacretaz-abanin,capizzi-eth-hydro,gen-hydro-nardis,gen-hydro-nardis,bulchandani-hydro-prb}.  We should also identify low-temperature states with large dynamical entropy, and compare quantitatively with OTOC's in large $N$ systems. Finally, we wish to {extend our work to monitoring multiple operators monitoring}~\cite{Gaspard1992-qchaos}, and propose a quantum many-body extension of the Pesin relation~\cite{Jorge,zurek-paz,decoherence-entropy-chaos,pesin-quantum}.


%
	
\begin{acknowledgments}
	I thank Denis Bernard, Jorge Kurchan and Tomaž Prosen for sharing their insight on related topics and encouraging me to pursue this work. I thank anonymous referees for valuable suggestions. 
	
\textit{Data availability.} The data that support the findings of this article are openly available \cite{data}.  
\end{acknowledgments}

\bibliography{refs}

\newpage
\section{End Matter}

\noindent\textit{Cluster decomposition and Gaussianity}
We argue for the Gaussianity of time-dependent fluctuations in noncritical systems. Such a system has a finite correlation length $\xi$ such that if $A_x$ and $A_y$ are operators has fixed support around $x$ and $y$, we have 
\begin{equation}
	\left< A_{x} A_y \right> =   \left< A_{x} \right> \left< A_y \right> + O(e^{-|x - y| / \xi}) \label{eq:clusterdecom}
\end{equation} 
as $|x - y| \gg \xi$.  Then, we claim that any $n$-time correlation function factorizes into two-point correlation functions, where the operator ordering is preserved. For example, for any fixed $t_1, \dots, t_4$, in the thermodynamic limit, we claim that
\begin{align}
	C_4 :=& \left< Q(t_1) Q(t_2) Q(t_3) Q(t_4) \right>  \nonumber \\
	=& C_{12} C_{34} + C_{14} C_{23} + C_{13} C_{24} +  O(1/\sqrt{V})   \label{eq:wick}
\end{align}
where $C_{ij} := \left< Q(t_i) Q(t_j) \right>$. To show \eqref{eq:wick}, first note that
\begin{equation} \label{eq:C4sum}
	C_4 = \sum_{x_1, \dots, x_4} \left< Q_{x_1}(t_1) \dots  Q_{x_4}(t_4) \right> / V^2 
\end{equation}
Now, $Q_{x_i}(t_i)$ is an operator centered around $x_i$, with a support size $v t_i$ modulo an exponential tail, where $v$ is the Lieb-Robinson velocity~\cite{LRbound}. Since we take the thermodynamic limit first, with $t_i$'s fixed, the linear system size $L = V^{1/d} \gg \xi' = \xi + v \max(t_i)$. 

Thus, the number of configurations with all points close to each other, $|x_i - x_j| \lesssim  \xi'$, are of order $V$, so their contribution is suppressed by a factor $V / V^2$, due to the factor $1 / V^2$ in \eqref{eq:C4sum}. 

For a configuration with a lonely point, e.g., with $|x_2 - x_{j}| \sim L$ for all $j \ne 2$, its contribution vanishes by \eqref{eq:clusterdecom} and the assumption $\left< Q_x(t) \right> = 0$. Indeed, 
\begin{align*}
&\left< Q_{x_1}(t_1) Q_{x_2}(t_2)\dots  Q_{x_4}(t_4) \right>   \\
=& \left< Q_{x_2}(t_2) Q_{x_1}(t_1) \dots  Q_{x_4}(t_4) \right>  \\
=& \left< Q_{x_2}(t_2) \right> \left< Q_{x_1}(t_1) \dots  Q_{x_4}(t_4) \right>  + O(e^{-L / \xi}) \to 0
\end{align*}
(since $x_1$ are $x_2$ are far, $ Q_{x_1}(t_1)$ and $Q_{x_2}(t_2)$ commute). 

We are left with contributions with two pairs, e.g., such that $|x_1 - x_3|, |x_2 - x_4| \lesssim \xi'$, but $ | x_1 - x_2| \sim L \gg \xi' $. Therefore by cluster decomposition, we have the factorization 
$  \left< Q_{x_1}(t_1) \dots  Q_{x_4}(t_4) \right> = \left< Q_{x_1}(t_1)  Q_{x_3}(t_3) \right> \left< Q_{x_2}(t_2)  Q_{x_4}(t_4) \right> + O(e^{- L  / \xi}) $.  Summing over the positions we obtain $C_{13} C_{24}$ up to vanishing corrections [one needs to invoke \eqref{eq:clusterdecom} for the two point functions]. Combined with the other two kinds of pairing configurations, we obtain  \eqref{eq:wick}.

The above argument generalizes to the $n$ point correlation $\left< Q(t_1) \dots Q(t_n) \right>$ and shows that it factorizes into two point functions, up to vanishing corrections. By continuation $t \to t + i \beta / 2$, the same ``Wick theorem'' holds for correlation functions across the thermofield double. Therefore, probed with a mesoscopic observable, with the thermodynamic limit taken first, a noncritical thermal state of is no different from a thermal state of a free boson model. This \textit{emergent} Gaussianity (or effective Gaussianity) property holds even if the microscopic model is interacting. As a consequence, we can compute dynamical entropy and purification rate with respect to  a mesoscopic observable and a Gaussian measurement scheme \eqref{eq:Km} in interacting systems analytically, by doing the computation in a free boson setup that has the same two-point functions.

\noindent \textit{Planck bound with FDT.} We may obtain a numerically tight bound on $\beta s_{\text{CNT}}$ given by the formula \eqref{eq:rateformula}, by maximizing that formula subject to the FDT constraint.  $ \beta s_{\text{CNT}}  \le  0.375\dots$. The optimal $G_K$ (see Fig.~\ref{fig:GKoptimal}) has a singularity $G_K \sim |\beta \omega|^{-\alpha}$ with $\alpha \approx 1$ at $\omega \to 0$. 

\begin{figure}
	\centering	\includegraphics[width=.9\columnwidth]{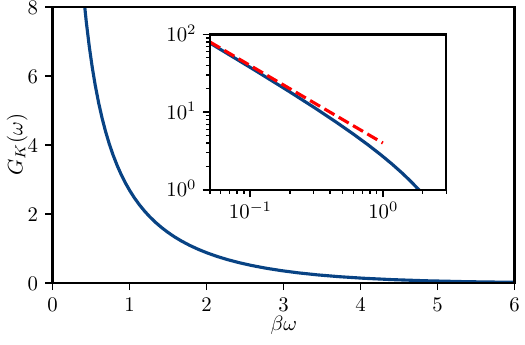}
	\caption{Keldysh correlation function $G_K$ (as function of $\beta \omega$) that maximizes $\beta s_{\text{CNT}}$ \eqref{eq:rateformula}. $G_K(-\omega) = G_K(\omega)$, obtained from numerically maximizing \eqref{eq:rateformula} subject to the fluctuation-dissipation theorem. Inset: comparison with $4 /  |\beta \omega |^{- 1}$ (dashed line). }\label{fig:GKoptimal}
\end{figure}

\end{document}